\documentclass[pdflatex,final,12pt]{elsarticle}
\usepackage[fontsize=11.5pt]{scrextend}
\journal{Decision Support Systems}

\usepackage{graphicx}
\usepackage{lineno,hyperref}
\usepackage{amsmath,amssymb,eucal,bm}
\usepackage{mathrsfs}
\usepackage{furumath}
\usepackage{version}	
\usepackage{color}

\newcommand{\team}{\text{(t)}}
\newcommand{\mem}{\text{(m)}}
\newcommand{\con}{\text{(c)}}
\newcommand{\out}{\text{(o)}}

\newcommand{\Oout}{\eO^\out}

\begin{document}

\title{Visual analytics of set data for knowledge discovery and member selection support}

\begin{frontmatter}
\author[Kyutech]{Ryuji Watanabe}
\ead{watanabe.ryuji717@mail.kyutech.jp}
\author[Kyutech]{Hideaki Ishibashi}
\ead{ishibashi@brain.kyutech.ac.jp}
\author[Kyutech]{Tetsuo Furukawa\corref{cor}}
\cortext[cor]{Corresponding author}
\ead{furukawa@brain.kyutech.ac.jp}
\address[Kyutech]{Kyushu Institute of Technology, 
2--4 Hibikino, Wakamatsu-ku, Kitakyushu 808-0196, Japan}

\begin{abstract}
Visual analytics (VA) is a visually assisted exploratory analysis approach in which knowledge discovery is executed interactively between the user and system in a human-centered manner. The purpose of this study is to develop a method for the VA of set data aimed at supporting knowledge discovery and member selection. A typical target application is a visual support system for team analysis and member selection, by which users can analyze past teams and examine candidate lineups for new teams. Because there are several difficulties, such as the combinatorial explosion problem, developing a VA system of set data is challenging. In this study, we first define the requirements that the target system should satisfy and clarify the accompanying challenges. Then we propose a method for the VA of set data, which satisfies the requirements. The key idea is to model the generation process of sets and their outputs using a manifold network model. The proposed method visualizes the relevant factors as a set of topographic maps on which various information is visualized. Furthermore, using the topographic maps as a bidirectional interface, users can indicate their targets of interest in the system on these maps. We demonstrate the proposed method by applying it to basketball teams, and compare with a benchmark system for outcome prediction and lineup reconstruction tasks. Because the method can be adapted to individual application cases by extending the network structure, it can be a general method by which practical systems can be built.
\end{abstract}
\end{frontmatter}

\renewcommand{\thefootnote}{}
\footnote{VA: visual analytics, TOI: target of interest, MNM: manifold network model, PVA: predictive VA, GPLVM: Gaussian process latent variable model, UKR: unsupervised kernel regression}
\renewcommand{\thefootnote}{\arabic{footnote}}
\setcounter{footnote}{0}

\section{Introduction}

Decision-making in member selection is difficult when we need to consider the combination effect of members. A typical example is member selection in team formation \cite{Moreno2012,Perez-Toledano2019a,Rahmanniyay2019}, where we need to consider the synergy between members \cite{Calder2015,Travassos2013,Araujo2016}. Another example is fashion outfit selection, where we need to select items, considering coordination \cite{Li2017}. The purpose of this study is to develop a method for supporting knowledge discovery and member selection using a visual analytics (VA) approach. 

VA is a visually assisted exploratory knowledge discovery process from a dataset, whereby users can explore data, examine hypotheses, gain knowledge, make predictions, and make decisions~\cite{Keim2008,Cui2019}. Such human-centered knowledge discovery is performed as an interaction between the user and VA system, where the system visualizes the analysis results according to the user's request, and the user feeds back its analysis intention, such as the target of interest (TOI) to the system. Through such an interactive loop, the VA system supports users in gaining a deeper insight into the data. The scope of VA is not limited to the analysis of past data, but also includes the prediction of new cases, whereby users can examine candidate options for decision-making.

The purpose of this study is to develop a general method for the VA of set data, which aims to support knowledge discovery from member set data, and support decision-making in member selection. A typical example is VA for sports teams (i.e., sets of athletes), whereby team managers can analyze the strength and weakness of past teams, as well as examine new lineups for future games. The VA approach is suitable for such management issues because it aims to support users in making decisions by themselves rather than discovering an optimal solution automatically on behalf of users. Thus, VA allows users to integrate their empirical domain knowledge with knowledge obtained from the system, and it leaves the final decision to users. Furthermore, the VA system also supports users not only in understanding {\em what} will happen as a result of the decision, but also in discovering {\em why} it will happen. This is in contrast to the conventional black-box approach, which only aims to discover an optimal solution automatically. 

Because there are several difficulties in handling set data, such as the combinatorial explosion problem, developing VA systems of set data is challenging. Thus, the goal of this study is to establish a general method for the VA of set data by which practical VA systems can be built, in particular, member selection support systems. The central issues of this study are (1) how the VA system should handle set data, (2) how the system should visualize the relation between relevant factors, such as the property of sets, their constituent elements, and output of the sets, (3) how the system should enable users to perform both exploratory analysis and prediction, (4) how the system should escape from the combinatorial explosion problem, and (5) how the system can be plastic so that it can be adapted to individual application cases. These five issues are the loci of interest of this paper. 

The key idea of this study is to model the generation process of sets and their outputs from low-dimensional latent variables using a manifold network model (MNM). Thus, to represent the generation process of sets, we connect manifold models of sets and of elements serially, whereas we connect manifold models of relevant factors in parallel, to represent the generation process of outputs. Using the MNM, sets, elements, and relevant factors are visualized on a set of latent spaces, similar to a set of topographic maps. These topographic maps are used as the interactive visual interface in which users can specify their TOI, and on which the corresponding information is visualized. This scheme is depicted in \figref{scheme}. The most important role of the target system is to help users to discover knowledge and hypotheses, and to consider team candidates through this interactive loop. 

The structure of this paper is as follows: In Section \ref{sec:background}, we introduce the background and related work. In Section \ref{sec:requirements}, we define the requirements of the target system. In Sections \ref{sec:framework} and \ref{sec:implementation}, we present the framework and implementation of the proposed method. In Section \ref{sec:results}, we demonstrate the proposed method by applying it to basketball team data. Finally, we present the discussion and conclusion in Sections \ref{sec:discussion} and \ref{sec:conclusion}. 

\begin{figure}
    \centering
    \includegraphics[width=0.8\linewidth]{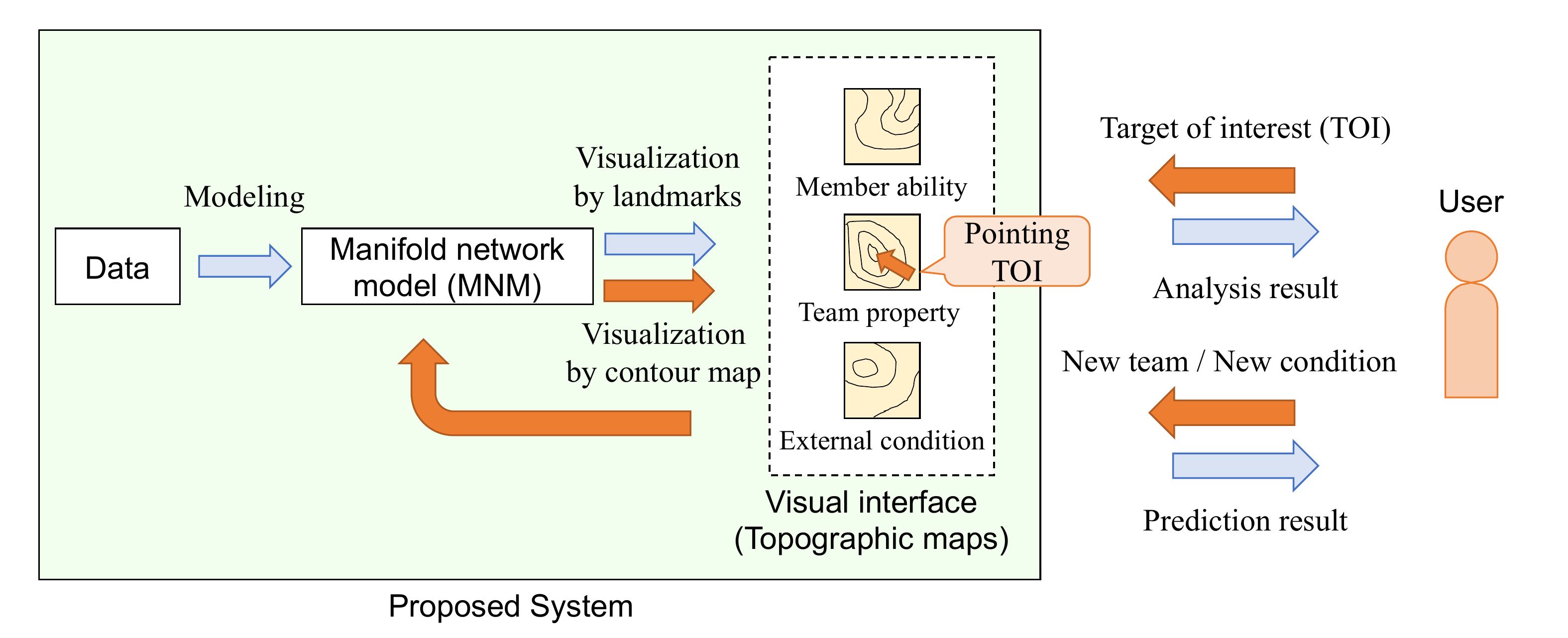}
    \caption{Key concept of the proposed method. The data are modeled as an MNM which is visualized on a set of topographic maps. In the topographic maps, the analysis targets (e.g., teams, constituent members, and external conditions) are visualized as a mapping, similar to landmarks, whereas other information (e.g., outcome) is visualized using coloring, similar to contour maps. Users can indicate their TOI on the topographic maps, and the system visualizes the corresponding information by changing the color.}
    \figlabel{scheme}
\end{figure}

\section{Background and related work} \label{sec:background}

\subsection{Visual analytics}

VA is a visually supported explanatory knowledge discovery process, or methods for the process. Keim et al.\ provided the definition of VA as follows: {\em ``Visual analytics combines automated analysis techniques with interactive visualizations for an effective understanding, reasoning and decision-making on the basis of very large and complex data sets''}~\cite{Keim2008}. The essential property of VA is that analysts themselves are regarded as an important part of the analytics process~\cite{Cui2019}. VA has been introduced to various fields, for example, telecommunications \cite{Wu2009}, supply chain management \cite{Park2016}, government \cite{Didimo2018}, education \cite{Qu2015}, and healthcare \cite{Caban2015}. 

VA aims not only at the exploratory analysis of past data, but also exploratory prediction for future cases. VA that particularly aims at prediction is referred to as predictive VA (PVA) \cite{Lu2017a}. In PVA, it is more important to help users to gain comprehensive knowledge of the prediction model rather than simply improving prediction accuracy. Therefore, in the decision-making process, users can obtain knowledge about {\em what} will happen and {\em why} it will happen as a result of the decision, and can integrate their empirical domain knowledge and management issues to make the final decision. This is in contrast to the conventional black-box approach, which focuses on prediction accuracy. The scope of this study includes PVA of sets. Thus, our target system allows users to examine new sets (e.g., new teams) by predicting their output.

\subsection{Handling a set of sets}

When handling set data, there are several serious problems. First, the treatment of the sets should be invariant to the permutation of the elements. Additionally, the cardinality (number of elements) is not always equal~\cite{Lee2019}. Second, the topological space of sets is not continuous, and there is no trivial definition of distance between two sets. This is a serious problem regarding achieving generalization for unknown sets. Third, the sets are often accompanied by combinatorial explosion problems.

In the case of discriminative tasks, our aim is to estimate a function $f$ from the given dataset, which represents the input-output relation as $y=f(X)$, where $X$ is a set. This $f$ should satisfy permutation invariance under arbitrary cardinality. Furthermore, $f$ should be generalized for unknown $X$, which is not included in the given dataset. Because there is no general method to obtain such an $f$, it is a challenging issue in the machine learning field~\cite{Zaheer2017,Lee2019}. 

By contrast, in the case of a generative task, our aim is to represent the probability of set $X$ as $p(X\mid \tau)$, where $\tau$ is the parameter that determines the property of the set. Again, the generative model $p(X\mid \tau)$ should satisfy permutation invariance. Although there is no established method, the most popular approach is to represent the generative model as $p(X\mid \tau)=\prod_i p(x_i\mid \tau)$, where $X=\{x_i\}$ is regarded as a set of independent and identically distributed random variables \cite{Edwards2017,Bouchacourt2017}. This approach is further categorized into two groups represented as 
$p(x\mid \tau) = \int p(x\mid\mu,\tau)\,p(\mu)\,d\mu$ \cite{Bouchacourt2017} and 
$p(x\mid \tau) = \int p(x\mid\mu)\, p(\mu\mid\tau) \,d\mu$ \cite{Edwards2017,Ishibashi2018}, where $\mu$ is the latent variable that determines the probability density of the element. In this study, we use the latter approach. 

For VA of set data, the system needs to visualize output $y$ of set $X$ (e.g., outcome $y$ of team $X$), in addition to generating or suggesting $X$ when the team property $\tau$ is specified. 
Thus, VA systems of set data need to equip three functions simultaneously: (i) predict the output for a given set ({\em discriminative task/forward problem}), (ii) generate new sets that satisfy the specified property ({\em generative task/inverse problem}), and (iii) visualize the property of various sets and their compositions ({\em visualization task}). Therefore, VA systems of sets suffer from difficulties in both discriminative and generative tasks, in addition to difficulties in visualization. Because equipping these three functions is a strict requirement, and they sometimes conflict with each other, these difficulties are the reason that few studies have been conducted on the VA of set data. 

\subsection{Member selection support for team management}

Decision-making in team formation and member selection is an important and difficult task; therefore, it is important to support it \cite{Mathieu2008,Bell2018}. 
In most studies, researchers have aimed to identify the optimal team lineup that maximizes the outcome by solving the combinatorial problem \cite{Moreno2012,Perez-Toledano2019a}. In this group of studies, the outcome was predicted by a function of the member lineup as $y=f(X)$, which was often defined a priori. Note that this approach only provides an optimal solution as a black-box system. In another group of studies, researchers aimed to evaluate members to determine or recommend suitable members for the team \cite{Calder2015,Uzochukwu2015,Jayanth2018}. Thus, they focused on individual members rather than a combination of members. 

Several studies on VA have been conducted in the team management field. Zhao et al.\ proposed a VA system for evaluating employee performance~\cite{Zhao2020}. Ryoo et al.\ applied VA to soccer players, which aimed to support player transfer between clubs \cite{Ryoo2018}. These studies were mainly focused on individual members rather than member combinations. Some other studies focused on the operation of teams, such as member collaboration \cite{Wu2019}. However, few studies have been conducted on VA for member selection that considers member combinations.

\section{Requirements of the target system} 
\label{sec:requirements}

In this section, we clarify the requirements that the VA system of set data should satisfy. Before considering general cases, we first imagine a case in which a team manager needs to determine members of a team. (For more practical cases, see \cite{PMBOK}.) To achieve this, the manager first needs to know what types of candidates the organization has. The manager needs to consider not only the abilities of individual candidates, but also their mutual influence, such as synergy. It would be also useful to know what types of teams have been formed and how many outcomes they have achieved. In some cases, the manager also needs to consider some extra factors, such as risks, costs, and external conditions. In the team sports case, the lineup of the opposing team is also an important factor. Considering these entangled factors, the manager needs to examine the candidate lineups, compare them, and then choose one of them. After the decision is made, the manager is also asked to account for the decision. 

Using the above scenario as a working assumption, we define the requisites that the target VA system should satisfy. First, the system needs to handle teams, that is, a set of members. More precisely, the system should be able to execute bidirectional mapping between teams and their properties. Thus, by specifying a lineup of a team, the system needs to estimate the property of the team, and estimate how many outcomes are expected (i.e., {\em forward problem}). Simultaneously, by specifying a property of a team, or specifying an expected outcome, the system needs to recommend lineups that satisfy the condition (i.e., {\em inverse problem}). Second, the system is required to unravel entangled factors, such as member ability, team property, other extra factors, and the outcome (i.e., {\em disentanglement problem}). Third, it is desirable that the system provides a table-top field where the manager can examine new lineups. It is also desirable to predict {\em what} will happen as a result of the decision and show {\em why} it will happen (i.e., {\em analysis} and {\em prediction}). Fourth, the system should free the manager from the individual consideration of candidate lineups by providing the manager with an overview of multiple lineups simultaneously at a glance (i.e., {\em visualization task}). 

These requirements are not limited to the team management case, but are generally necessary for the VA system of set data. The above requirements are translated as follows: (1) Clearly, the VA system of set data should be able to handle sets. Thus, the system should be capable of bidirectional (i.e., forward and inverse) mapping between a given set and its property. (2) The VA system of set data should support users in both intra-domain and inter-domain analysis. Particularly, the system is required to unravel the entangled relation of relevant factors using inter-domain visualization. (3) The system should be able to support both the analysis of past data and prediction of future cases. For this purpose, the system needs to obtain the predictive model from past data. (4) The system is required to eliminate the combinatorial explosion problem from which users suffer. Thus, it is desirable for the system to visualize the output of various sets simultaneously to provide users with an overview of them at a glance. 

Additionally, we add an extra requirement for the development method: the method should allow us to design VA systems flexibly so that they can adapt to individual cases. Because the input (available dataset) and output (visualized aspects that the user wants to see) are different for each case, even if they are in the same application field, such generality is required for the development method. Additionally, the VA process itself is dynamic~\cite{Cui2019}; after obtaining some insights or hypotheses, it is often necessary for the user to perform further analysis from a different viewpoint, by adding a new dataset, or by visualizing additional aspects. Therefore, it is desirable for VA systems to be both adaptable and expandable. In this study, this is referred to as the {\em plasticity requirement}. 

Our aim in this study is to establish a general method for constructing VA systems of set data that satisfies these five requirements. To the best of our knowledge, no studies have tackled these issues, presumably because of the difficulties of handling sets in VA systems.

\section{Framework of the proposed method} \label{sec:framework}

\subsection{Manifold assumption} \label{subsec:assumption}

For the system to escape from the combinatorial explosion problem, the exploration space must be restricted. Additionally, the system needs to estimate the predictive model from the limited number of past data, which do not cover the possible combinations exhaustively. Therefore, we use the manifold assumption in this study, which is common in the VA field \cite{Cui2019}. Thus, we postulate that meaningful team compositions are distributed in low-dimensional space. More properly, we suppose a continuous topological space of team compositions $\eP$ in which the distance between two teams is defined. Then we assume that the team compositions that are worth considering are in a low-dimensional manifold in $\eS^\team\subseteq\eP$, whereas all other compositions are not of interest. This assumption allows us to restrict the exploration space within the manifold, and allows us to visualize team compositions more easily. Additionally, the assumption also means that team compositions that were observed in the past are distributed in the manifold, whereby the system becomes able to estimate the predictive model from a non-exhaustive dataset. Because two-dimensional space is convenient for visualization, we further assume that we are interested in the two-dimensional principal dimensions of the manifold. Under this assumption, we can define a bidirectional mapping between $\eS^\team$ and the latent space $\eL^\team$, which is used for visualization. Thus, we can define a bijection $\pi\colon\eL^\team\leftrightarrow\eS^\team$, whereby a team at $\tau\in\eL^\team$ is mapped to $\eS^\team\ni q_\tau=\pi(\tau)$, and vice versa. This bijection $\pi$ plays an essential role in bidirectional exploration in the VA system.  

Similarly, we also assume that member properties are distributed in low (typically two)-dimensional manifolds in their feature space. Thus, we assume that the member feature vectors are distributed in manifold $\eS^\mem\subseteq\eO^\mem$, where $\eO^\mem$ is the vector space of member features. Then we can define the homeomorphic latent space $\eL^\mem$, and a bidirectional mapping $g\colon\eL^\mem\leftrightarrow\eS^\mem$, whereby a member at $\mu\in\eL^\mem$ is mapped to the feature space as $\vx=g(\mu)$, and vice versa. Similarly, we assume the manifolds of extra factors. For example, when we need to consider an external condition, manifold $\eS^\con\subseteq\eO^\con$ and the latent space $\eL^\con$ can be defined, in addition to the bijection $h\colon\eL^\con\leftrightarrow\eS^\con$. 

The manifold assumption results in another benefit: it allows us to represent the member composition of a team as a probability density in the member latent space $\eL^\mem$ as $p(\mu)$. Thus, the team composition space $\eP$ is defined as the function space that consists of the probability densities on $\eL^\mem$. Therefore, in our method, teams are treated as probability densities instead of discrete sets. This approach allows us to measure the distance between two teams and generate new lineups. This is the key idea for handling sets in the proposed method.

\subsection{Framework of the proposed method}

\begin{figure}[t]
 \centering
   \includegraphics[width=0.9\linewidth]{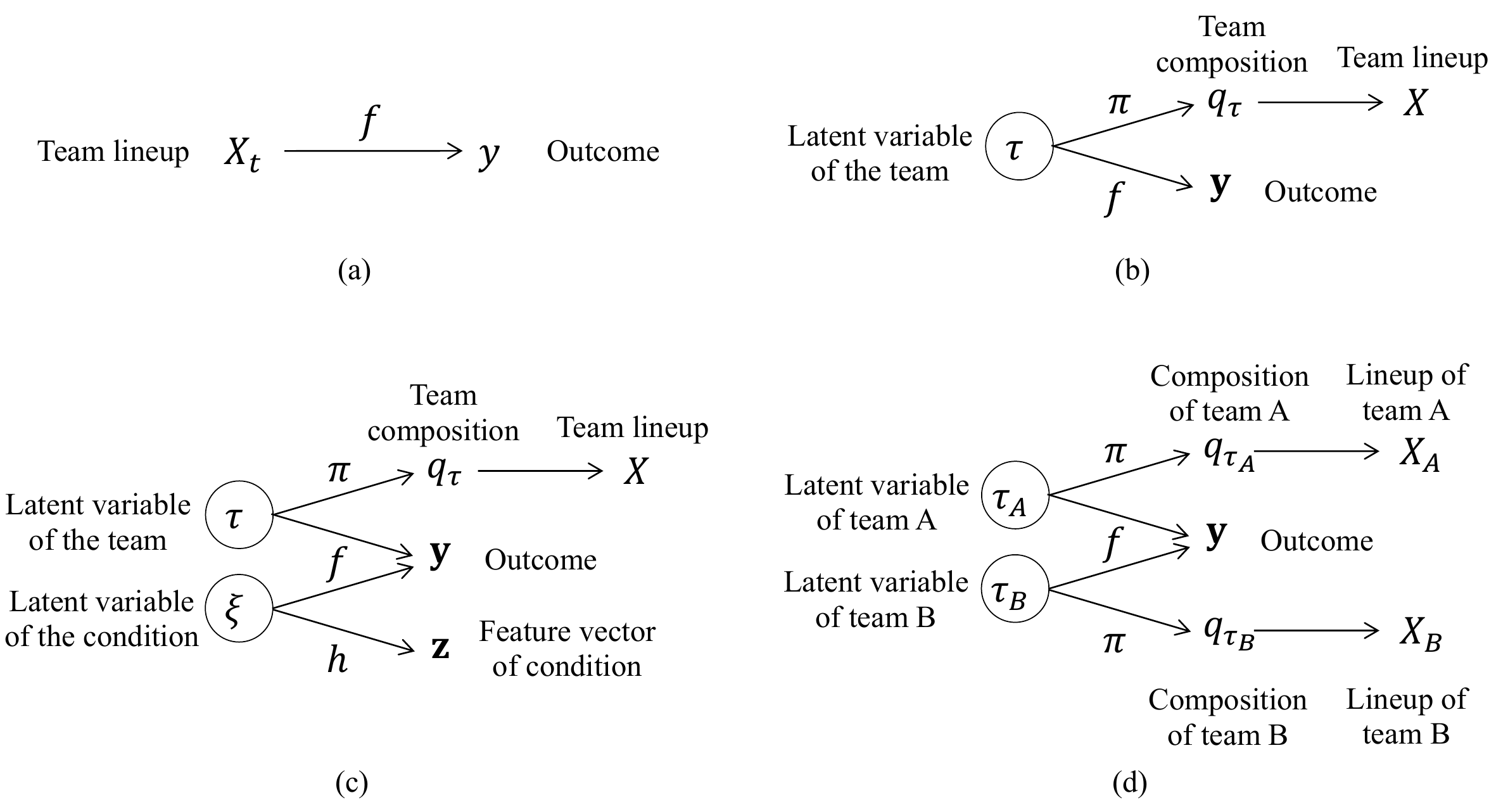}
  \caption{Conceptual diagrams of the traditional (a) and proposed (b, c, d) approaches. Circles represent the latent variables. (a) Traditional optimization approach. (b) Generative model approach using the team latent variable $\tau$. (c) Generative model approach with external conditions. (d) Generative model of sport teams in a match.}
\figlabel{GenerativeModel1}
\end{figure}
\begin{figure}[t]
 \centering
 \includegraphics[width=0.48\linewidth]{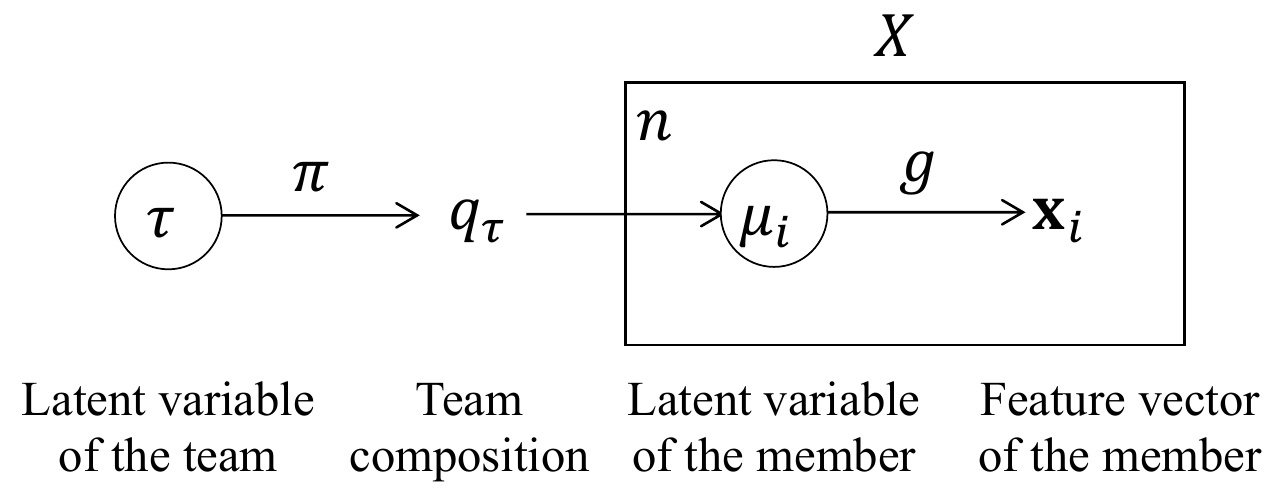}
\caption{Generative model of team lineup $X$ from the team latent variable $\tau$. The variables indicated by circles are latent, and the box indicates that the generative process repeats $n$ times.}
\figlabel{GenerativeModel2}
\end{figure}

Now we formulate the framework of the proposed approach. Let $\Omega$ be the total set of members. A team that consists of $n$ members is represented as a set $T=\{\omega_1,\dots,\omega_n\}$, where  $\omega_i\in\Omega$. In this paper, $T$ is referred to as the {\em lineup} of the team. Let $\mathfrak{T}$ be the set of all lineups\footnote{In this paper, scalars and vectors are indicated by lowercase italic and bold face, respectively (e.g., $x$ and $\vx$), whereas matrices are indicated by uppercase boldface (e.g., $\vX$). Sets are indicated by uppercase italic (e.g., $T)$, whereas a set of sets is indicated by Fraktur letters (e.g., $\mathfrak{T})$. The upper limit of integer indices are also indicated by uppercase italic (e.g., $N$). Continuous spaces and manifolds are represented in cursive script (e.g., $\eL$, $\eS$). The superscripts (t), (m), (c), and (o) denote {\em team, member, condition, and outcome}, respectively (e.g., $\eL^\team$, $\eO^\mem$).}. Thus $\mathfrak{T}$ is a set of sets. Note that the order of the members is not fixed generally. Let $\vx(\omega)\in\eO^\mem$ be the feature vector of member $\omega$, which represents the property of the member, such as ability. Then we have a set of feature vectors with respect to lineup $T$: $X_T=\{\vx(\omega_1),\dots,\vx(\omega_n)\}$. Because $T$ and $X_T$ are often identified, we also refer to $X_T$ as the lineup. 

In a conventional framework, we typically assume that the outcome $y$ is determined as a function of $X_T$, say $y=f(X_T)$. Simply speaking, a typical traditional task is to identify the optimal lineup $T_\text{opt}\in\mathfrak{T}$ that maximizes the outcome $y$ (\figref{GenerativeModel1} (a)). Note that outcome $y$ should be scalar in this approach. Additionally, $f$ is typically given or designed by authors. 

By contrast, the proposed framework assumes that both lineup $X$ and outcome $\vy$ (which can be vectors in our case) are generated from the latent variable $\tau\in\eL^\team$, which represents the intrinsic property of the team (\figref{GenerativeModel1} (b)). Thus, the outcome is generated as $\vy=f(\tau)$ from $\tau$ instead of from lineup $X$ directly. Unlike the conventional framework, $f$ is not given a priori, but is estimated in a data-driven manner. Note that outcome $\vy\in\eO^\out$ is distributed in a manifold $\eS^\out\subseteq\eO^\out$. Parallel to the outcome, $\tau$ also generates a probability density $q(\mu\mid\tau)\equiv q_\tau(\mu)$, which represents the member composition of the team. Thus, the generative model of team lineups is represented as
\begin{align}
    q(X\mid\tau) &= \prod_i \int \N{\vx_i}{g(\mu),\beta\inv\vI}\,q_\tau(\mu)\,d\mu,
\end{align}
where $q_\tau=\pi(\tau)$, and $\beta\inv$ is the variance ($\vI$ is the unit matrix)  (\figref{GenerativeModel2}). Conversely, when a lineup $X=\{\vx_1,\dots,\vx_n\}$ is given, the corresponding latent variable $\tau(X)$ is determined as
\begin{align}
\tau(X) &= \argmin_\tau \DKL{p(\mu\mid X)}{q_\tau(\mu)},
\end{align}
where $p(\mu\mid X)$ is given by the kernel density estimator, and $D_\text{KL}$ is the Kullback--Leibler (KL) divergence. Such bidirectional mapping allows bidirectional exploration between the team latent variable and lineups, and it is an advantage of this method.

\begin{figure*}[t]
    \centering
    \includegraphics[width=0.8\textwidth]{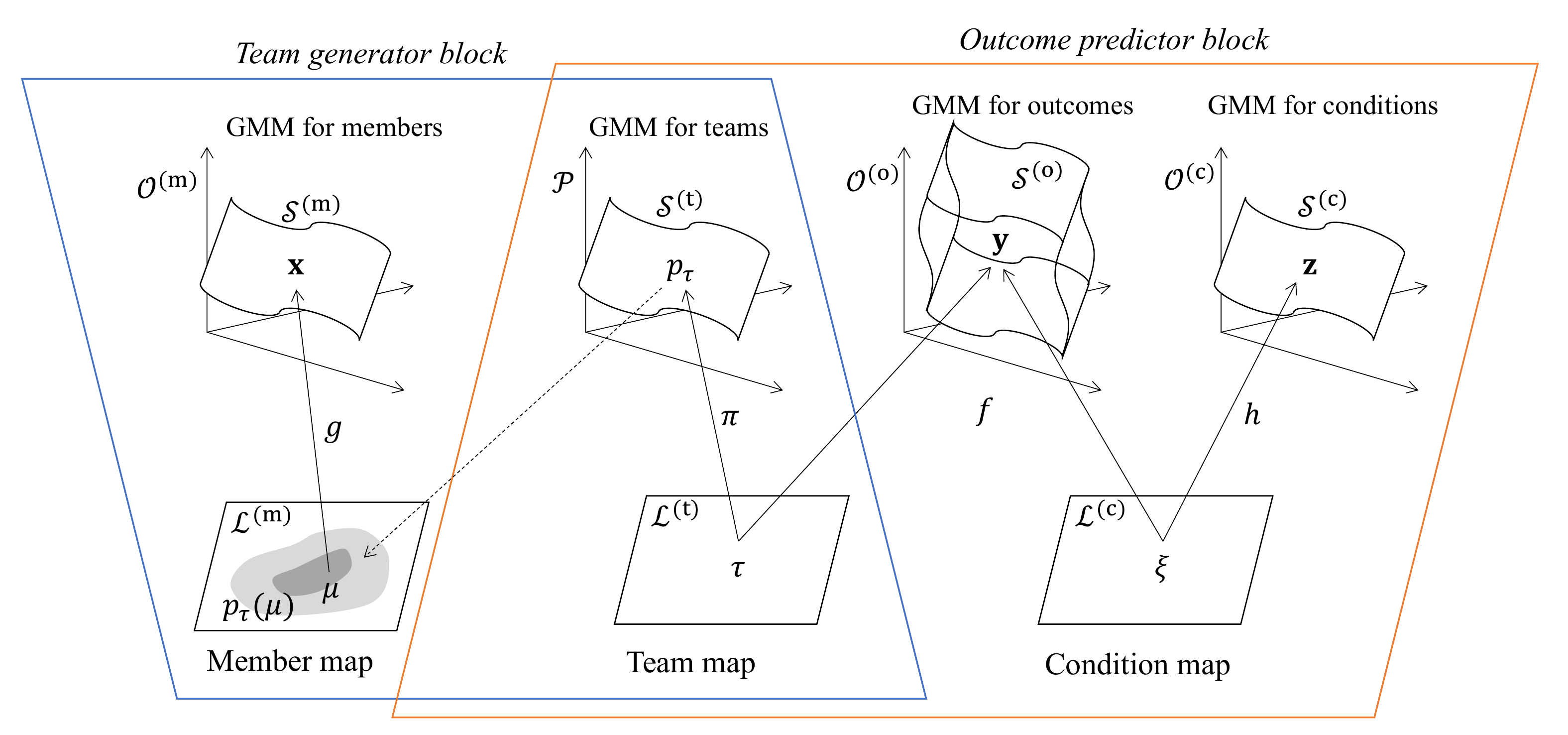}
    \caption{Generative model and the structure of the proposed method for the scenario of \figref{GenerativeModel1} (c). The MNM consisting of four manifold models corresponding to the four outputs, and consisting of three topographic maps corresponding to the three latent spaces.}
    \figlabel{structure}
\end{figure*}

This generative model can easily adapt to any scenario in which some additional factors are to be considered. If we need to consider an external condition, then the model is extended by adding another latent variable $\xi\in\eL^\con$, which represents the intrinsic property of the condition. In this case, the expected outcome is predicted as $\vy=f(\tau,\xi)$ (\figref{GenerativeModel1} (c)). Another typical case is sport teams in a match, where victory or defeat depends on the latent variables of both teams, such as $\vy=f(\tau_A,\tau_B)$ (\figref{GenerativeModel1} (d)). 

In our approach, the entire generative model becomes a network of manifold models, that is, the MNM{}. \figref{structure} shows the network structure of the MNM, when an external condition is considered (\figref{GenerativeModel1} (c)). In the MNM, the generation process of lineups is represented by a serial connection of two manifold models $\eS^\team$ and $\eS^\mem$ via latent space $\eL^\mem$, whereas the outcomes and relevant factors are modeled by the parallel connection of manifolds $\eS^\team$, $\eS^\con$, and $\eS^\out$ via latent spaces $\eL^\team$ and $\eL^\con$. Because the network structure can be adapted to individual application cases, the proposed framework satisfies the plasticity requirement. 

\subsection{Definition of the learning task} \label{sec:LearningTask}

In our framework, the manifold models are estimated in a data-driven manner. In the following, we clarify what the method needs to estimate from the given dataset when external conditions are considered (\figref{structure}). Suppose that the given dataset consists of $N^\team$ lineups $\mathfrak{L}=\left\{T_i\right\}_{i=1}^{N^\team}$, with $N^\mem$ members $\vX=\left(\vx_j\right)_{j=1}^{N^\mem}$, under $N^\con$ external conditions $\vZ=\left(\vz_k\right)_{k=1}^{N^\con}$, where $T_i\in\mathfrak{T}$, $\vx_j\in\eO^\mem$, and $\vz_k\in\eO^\con$. Furthermore, let $\vY=\left(\vy_n\right)_{n=1}^{N^\out}$, $\vy_n\in\Oout$ be the outcome data observed from the combinations of $\left\{(T_{i(n)},\vz_{k(n)})\right\}_{n=1}^{N^\out}$. Note that it is not necessary to cover all combinations, and missing combinations are allowed. Under such a dataset, the task of the proposed method is to estimate the mappings $f$, $g$, $h$, and $\pi$, in addition to estimating the latent variables $\vT=\left(\tau_i\right)_{i=1}^{N^\team}$, $\vM=\left(\mu_j\right)_{j=1}^{N^\mem}$, and $\vXi=\left(\xi_k\right)_{k=1}^{N^\con}$.

\section{Implementation} \label{sec:implementation}
\subsection{Generative manifold modeling} 

As described above, the core idea of our proposed approach is to represent the generative model using the MNM{}. To achieve this, we adopt generative manifold modeling (GMM) as a building block in the implementation. Thus, the MNM is estimated by the network of GMM{}. This is the key idea of the implementation. 

The term GMM in this paper refers to the paradigm of unsupervised learning, which aims to model the given dataset using a manifold. For this purpose, GMM estimates a map from low-dimensional latent space to high-dimensional data space, in addition to estimating the low-dimensional latent variable for each sample. After learning is complete, GMM allows users to conduct both retro- and anterograde surveys, that is, analysis and prediction. This is why we use GMM in our method. The representatives of GMM are the Gaussian process latent variable model (GPLVM) \cite{Lawrence2005a}, and unsupervised kernel regression (UKR) \cite{Meinicke2005}.

Let $\eO=\bbR^D$ and $\eL=\bbR^d$ be high-dimensional observable space and low-dimensional latent space, respectively. For a high-dimensional dataset $\vX=(\vx_1,\dots,\vx_N)^\top\in\bbR^{N\times D}$, the task of GMM is to estimate the corresponding latent variables $\vXi=(\xi_1,\dots,\xi_N)\in\bbR^{N\times d}$, and represent the mapping $f\colon\eL\to\eO$ as $\vx=f(\xi\mid\vXi)$. For this purpose, GPLVM uses the Gram matrices of observed data $\vS=\vX\vX^\top$ and of latent variables $\vK=\left(k_{ij}\right)$, where $k_{ij}\equiv k(\zeta_i,\zeta_j)$ is the kernel function. The objective function of GPLVM is defined as
\begin{align}
F_\text{GPLVM}(\vXi\mid\vX)
   &= -\frac{ND}{2}\ln 2\pi - \frac{D}{2}\ln |\Hat\vK|
      -\frac12\Tr{\Hat\vK\inv \vS},
\end{align}
and (the mean of) the mapping is determined by $f(\xi\mid\vXi) = \vk(\xi)^\top \Hat\vK\inv \vX$, where $\Hat\vK= \vK+\beta\inv\vI$ and $\vk(\xi) = \left(k(\xi,\xi_i)\right)_{i=1}^N$, and $\beta\inv$ is the variance. An advantage of using GPLVM is that we obtain not only the average mapping $f(\xi\mid\vXi)$, but also its variance $\sigma(\xi,\xi')$, which represents uncertainty. By contrast, the objective function of UKR is defined as
\begin{align}
F_\text{UKR}(\vXi|\vX)
  &= -\frac\beta2 \sum_i \norm{f(\xi_i)-\vx_i}^2, 
  \eqlabel{OF_UKR}
\end{align}
and the mapping is determined as $f(\xi\mid\vXi) = \frac1{K(\xi)}\sum_{j=1}^N k(\xi,\xi_j)\, \vx_j$, where $K(\xi) = \sum_{j=1}^N k(\xi,\xi_j)$. An advantage of UKR is that it can deal with a set of probability distributions as a dataset. In this case, the Euclidean distance in \eqref{OF_UKR} is replaced by the KL divergence. We use GPLVM for modeling the outcome predictor $f$, whereas we use UKR for modeling the team composition generator $\pi$.

\subsection{Estimation of the manifold network model using GMM}

The structure of the MNM consists of four types of connections: (a) mapping from a latent space to the counterpart manifold (e.g., mapping $g$ from $\eL^\mem$ to $\eS^\mem$ in \figref{structure}); (b) serial connection between the manifold of sets and the latent space of elements (i.e., the link between $\eL^\mem$ and $\eS^\team$); (c) mappings from a latent space to two (or more) manifolds (e.g., mappings $\pi$ and $f$, from $\eL^\team$ to $\eS^\team$ and $\eS^\out$); and (d) mappings from two (or more) latent spaces to a manifold (e.g., mapping $f$, from $\eL^\team$ and $\eL^\con$ to $\eS^\out$). These links are estimated by GMM as follows:

\begin{description}
  \item[(a)] To estimate a mapping from a latent space to the counterpart manifold, ordinary GMM (either GPLVM or UKR) can be used. 
  \item[(b)] To transform a team composition from team manifold $\eS^\team$ to the lineup in member latent space $\eL^\mem$, member latent variables are generated by the team composition $q(\mu\mid\tau)$. By contrast, from $\eL^\mem$ to $\eS^\team$, a team lineup $X=\{\vx_1,\dots,\vx_n\}$ is transformed to the team composition as $p_X(\mu)=\frac1n\sum_i k(\mu\mid g\inv(\vx_i))$ using the kernel density estimator. If the lineup is defined by a weighted set, then it becomes $p_X(\mu)=\frac1W\sum_i w_i k(\mu\mid g\inv(\vx_i))$, where $w_i$ is the weight of the $i$-th member, and $W=\sum_i w_i$. For example, in the sports game case, weight $w_i$ can be defined by the playing time of the $i$-th athlete.
  \item[(c)] To estimate mappings from a latent space to two manifolds, two objective functions of GMM are summed with the weight coefficient as $\alpha_1 F_1+\alpha_2 F_2$. In this study, the coefficients $\alpha_i$ are determined so that the average norms of the gradient vectors are approximately equal. 
  \item[(d)] To estimate a mapping from two latent spaces to a product manifold, the joint kernel function is defined as the product of two kernels. Thus, the kernel function for $f$ is defined as $k(\tau,\xi,\tau',\xi')\equiv k(\tau,\tau')\times k(\xi,\xi')$. 
\end{description}
In the case of \figref{structure}, the objective functions of four GMM blocks are integrated into the following two objective functions: 
\begin{align}
 F(\vM) &= F^\mem(\vM\mid\vX),\\
 F(\vT,\vXi) &= \alpha_1 F^\team(\vT\mid P)
    + \alpha_2 F^\con(\vXi\mid\vZ)
    + \alpha_3 F^\out(\vT,\vXi\mid\vY), 
\end{align}
where $P=\{p_i(\mu)\}$ is the set of team compositions. 

Regardless of the network structure, any MNM can be estimated by the combination of these four styles of GMM connections. Therefore, the plasticity requirement is also satisfied at the implementation level.

\section{Demonstration of the proposed VA system} \label{sec:results}

\begin{figure*}
  \centering
  \includegraphics[width=0.9\textwidth]{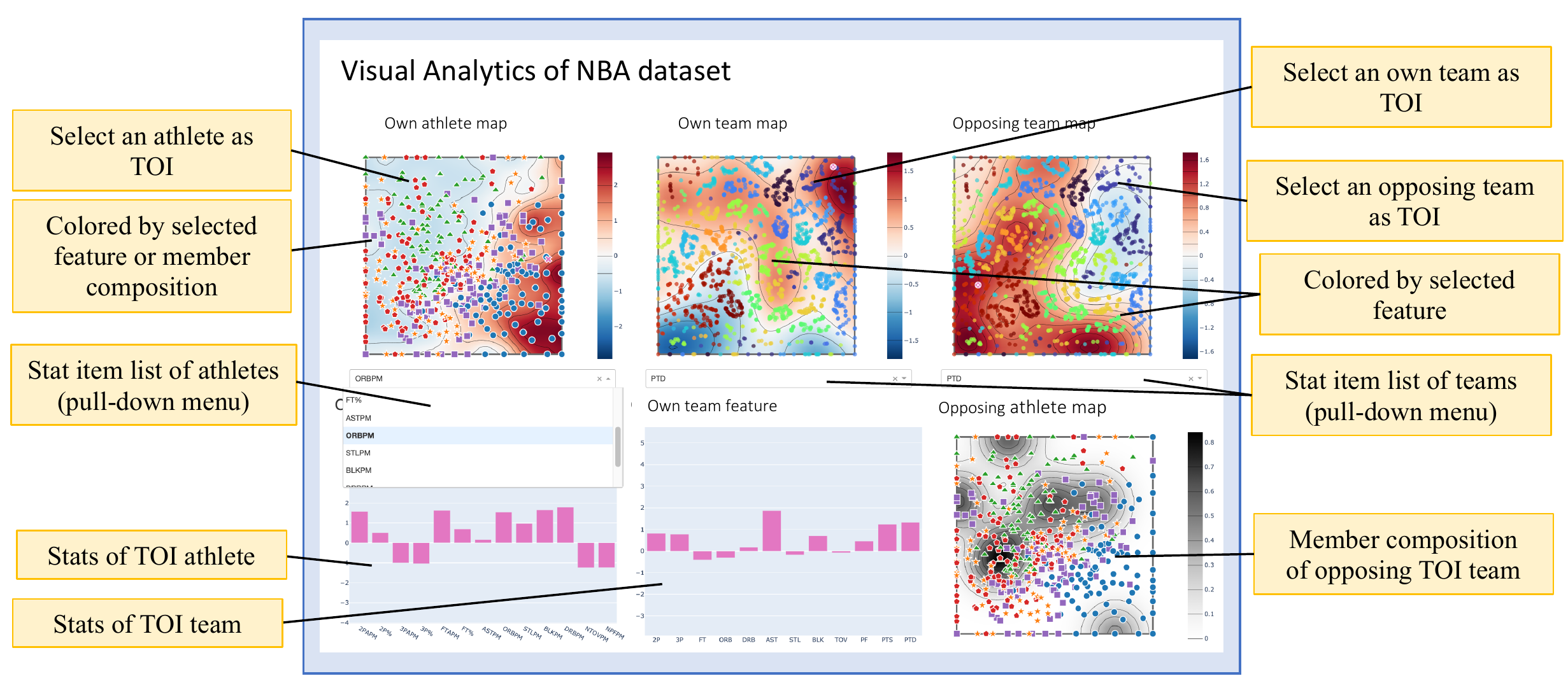}
  \caption{Visual analytics system for the NBA dataset.}
  \figlabel{NBA-system}
\end{figure*}

In this section, we demonstrate the proposed method by applying it to real team data, and show how the interactive analytics process can be executed using our system. We used basket team data from the National Basketball Association (NBA)\footnote{The details of the NBA dataset are as follows: We used the data of 2,456 lineups from 1,228 games held in 2018--2019 obtained from the Basketball Reference website \url{https://www.basketball-reference.com/}. We used the data of 1,105 out of 1,228 games for training, and the remaining 123 games for validation and testing. We used 12 statistics of teams, such as point difference (PTD) and number of offensive rebounds, which are regarded as the outcome. We also used the data of 13 statistics of 492 NBA athletes, such as two-point goals attempted (see the website for the details.) In this experiment, we regarded these 492 NBA athletes as the entire population, and assumed that teams were to be constructed by the selection of members from the population. Neither the basketball club nor position information of athletes were used in this demonstration. We regarded lineups as weighted sets, in which weights were determined by the playing time in the game.}. The developed VA system is shown in \figref{NBA-system}, which is based on the generative model shown in \figref{GenerativeModel1} (d). In the proposed method, the latent spaces are visualized as square spaces, such as a set of topographic maps. In the case of the NBA dataset, the system has four topographic maps: {\em own and opposing athlete maps}, and {\em own and opposing team maps}. The own and opposing maps are essentially identical, and they can be colored differently according to the TOI of users. In addition to these topographic maps, there are two bar charts that display the statistics of the athlete and team (i.e., the outcome), and pull-down menus that enable the user to select a stat to be visualized. 

\begin{figure*}[t]
    \centering
    \includegraphics[width=0.8\textwidth]{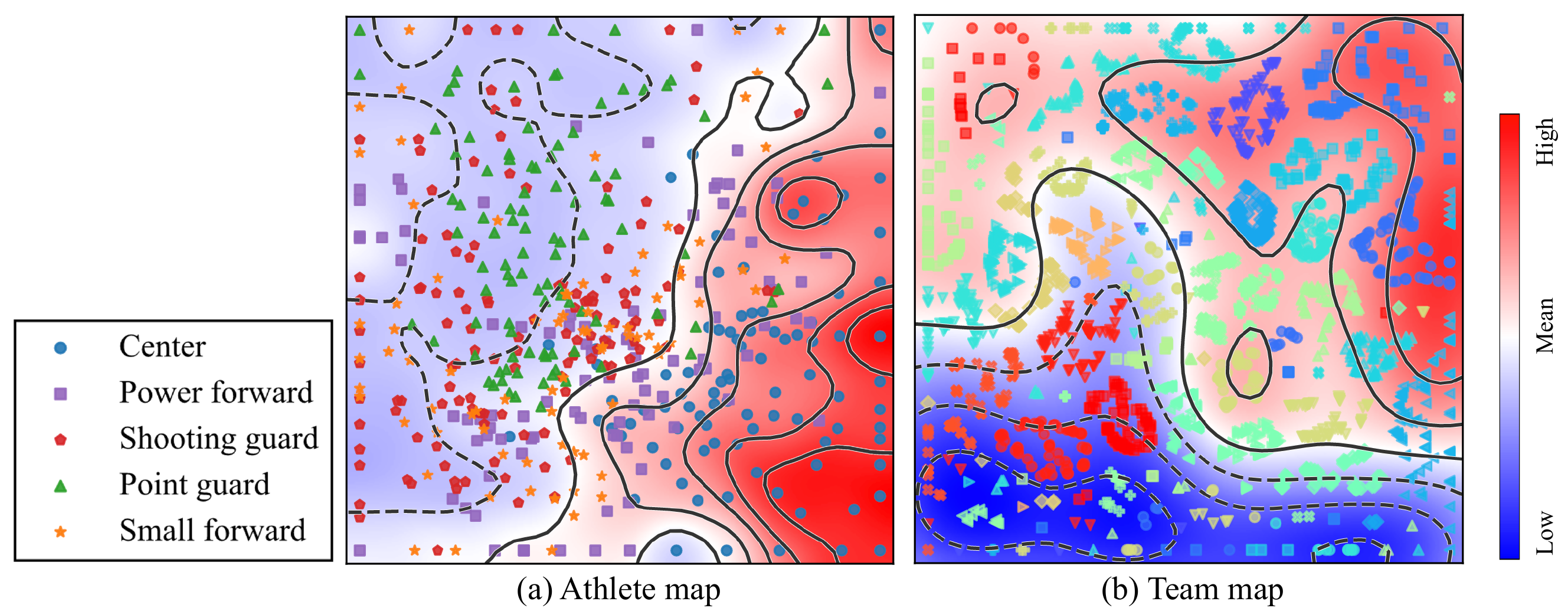}
    \caption{Visualization results of athletes and teams for the NBA dataset.  (a) Athlete map: markers indicate the athletes and their positions, and the color scale indicates the number of offensive rebounds per minute. (b) Team map: markers indicate the teams and the clubs to which they belong. The color scale shows the mean of the PTD.} 
    \figlabel{result1}
\end{figure*}

\begin{figure*}[t]
    \centering
    \includegraphics[width=0.9\textwidth]{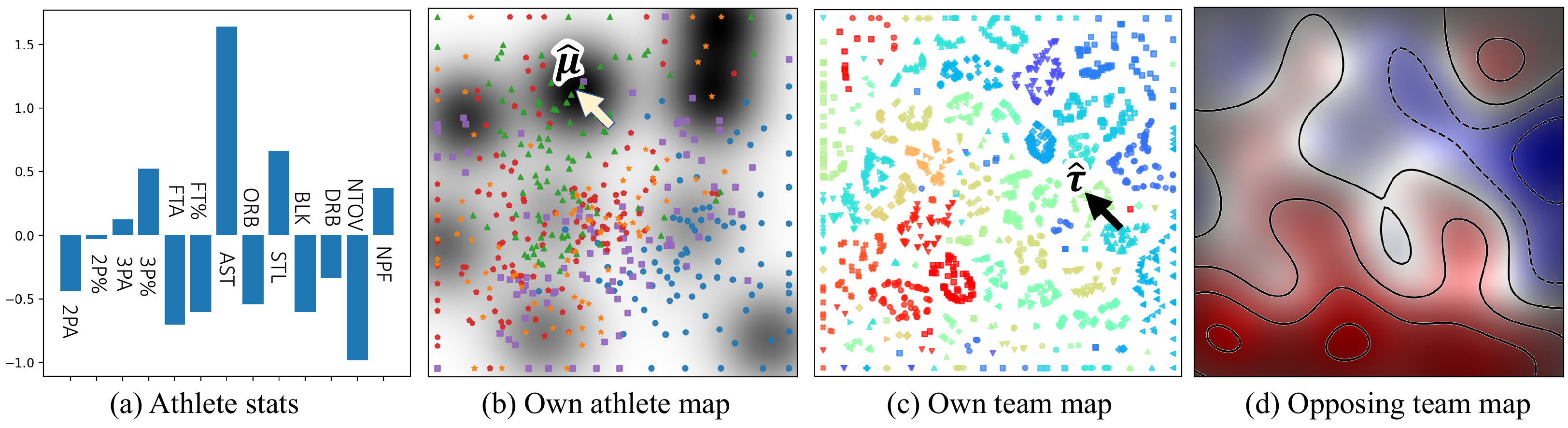}
    \caption{Example of interactive visualization. The TOI is team $\Hat\tau$ indicated on the own team map (c) by the user. The predicted PTD is shown on the opposing team map (d), which indicates that team $\Hat\tau$ would defeat the opposing teams in the red region. Simultaneously, the member composition of the team $p(\mu\mid\Hat\tau)$ is visualized on the athlete map (b). If the user selects one of the members ($\Hat\mu$), the stats of the member are indicated as a bar chart (a). }
    \figlabel{result2}
\end{figure*}

\figref{result1} shows the obtained athlete map and team map. In the athlete map, athletes are indicated as fixed points, like landmarks, so that athletes who have similar stats are arranged close to each other. This can be seen from the fact that the athletes are roughly divided according to their positions on the map. Similarly, teams are visualized as landmarks on the team map, in which teams that have similar stats are arranged close to each other. 

In the case of real topographic maps, we can visualize additional information, such as altitudes and population density, using a contour map or heat map. In the same way, the topographic maps in our system can be colored according to one of the stats. In \figref{result1} (a), the athlete map is colored according to the offensive rebound per minute. When a user wants to see another stat (e.g., number of two-point goals attempted), the user selects the stat from the pull-down menu. In the same way, the team map (\figref{result1} (b)) is colored according to the average point difference (PTD), which indicates the chance of winning (higher for the red region). If the user selects a stat from the pull-down menu, the system shows other stats on the map, which enables the user to analyze the strength and weakness of the teams.  

These topographic maps are used as the bidirectional visual interface by which the user can indicate its TOI in the system. \figref{result2} shows an example of interactive analysis. Suppose that the TOI of the user is the team located at $\Hat\tau$ on the own team map (c). If the user points to $\Hat\tau$ on the map, the system visualizes the corresponding member composition $q(\mu\mid\Hat\tau)$ in grayscale on the athlete map (b), which shows the population density of team $\Hat\tau$ on the athlete map. The user can further see the stats of each constituent member as a bar chart by pointing to it on the athlete map. Simultaneously, the system also visualizes the predicted game result such as the PTD on the opposing team map (d). Thus, the winning chance for team $\Hat\tau$ is higher when it plays a game against the opposing teams indicated in the red region. This is achieved by visualizing $f(\Hat\tau,\tau_\text{opp})$ as the function of $\tau_\text{opp}$. Additionally, the confidence of the prediction can be indicated by the brightness of the color by visualizing $\sigma(\Hat\tau,\tau_\text{opp})$. 

As shown in the above example, the proposed system solves the forward and inverse problems seamlessly. We evaluated the accuracy of the forward problem using game result (i.e., winning/losing) prediction, and the system yielded 63\% accuracy\footnote{In previous studies, the prediction accuracy ranged from 60\% to 80\% \cite{Horvat2020}, and our result was within this range. Note that accuracy was evaluated on different datasets and under different task constraints; hence, a fair comparison was difficult.}, where the baseline was 50\% (i.e., the chance level). For comparison, we constructed a ``benchmark system'' based on the average stats of members\footnote{For a fair comparison, we carefully designed the benchmark system as follows: We transformed each lineup into a feature vector using the average stats of members. Then we mapped the feature vector to two-dimensional space using principal component analysis (PCA). This two-dimensional vector corresponded to the team latent variable $\tau_i$. Finally, we trained the predictor using Gaussian process regression. Note that the benchmark system does not fulfill the requirements of the VA system because it cannot solve the inverse problem.}, which yielded 56\% accuracy. We also evaluated the accuracy of the inverse problem using the reconstruction error of member compositions. Thus, we evaluated the Jensen--Shannon (JS) divergence \cite{Lin1991} between the test member composition $p^\text{test}_i(\mu)$ and the composition reconstructed from the model $q(\mu|\tau^\text{test}_i)$ for each test lineup. The reconstruction error of the proposed method was $0.012$, whereas the baseline was $0.24$ (measured using the uniform distribution). For comparison, we also measured the JS divergence between the test lineups and the average member composition $\ol{p} (\mu)$, where $\ol{p}(\mu)=\frac1{N^\team} \sum_i p_i^\text{training}(\mu)$. It was $0.14$, which means that the degree of lineup variety was much larger than the reconstruction error of the proposed system. Note that the reconstruction error of member compositions cannot be measured for methods that do not consider lineup sets, including the benchmark system.

\begin{figure*}[t]
    \centering
    \includegraphics[width=0.9\textwidth]{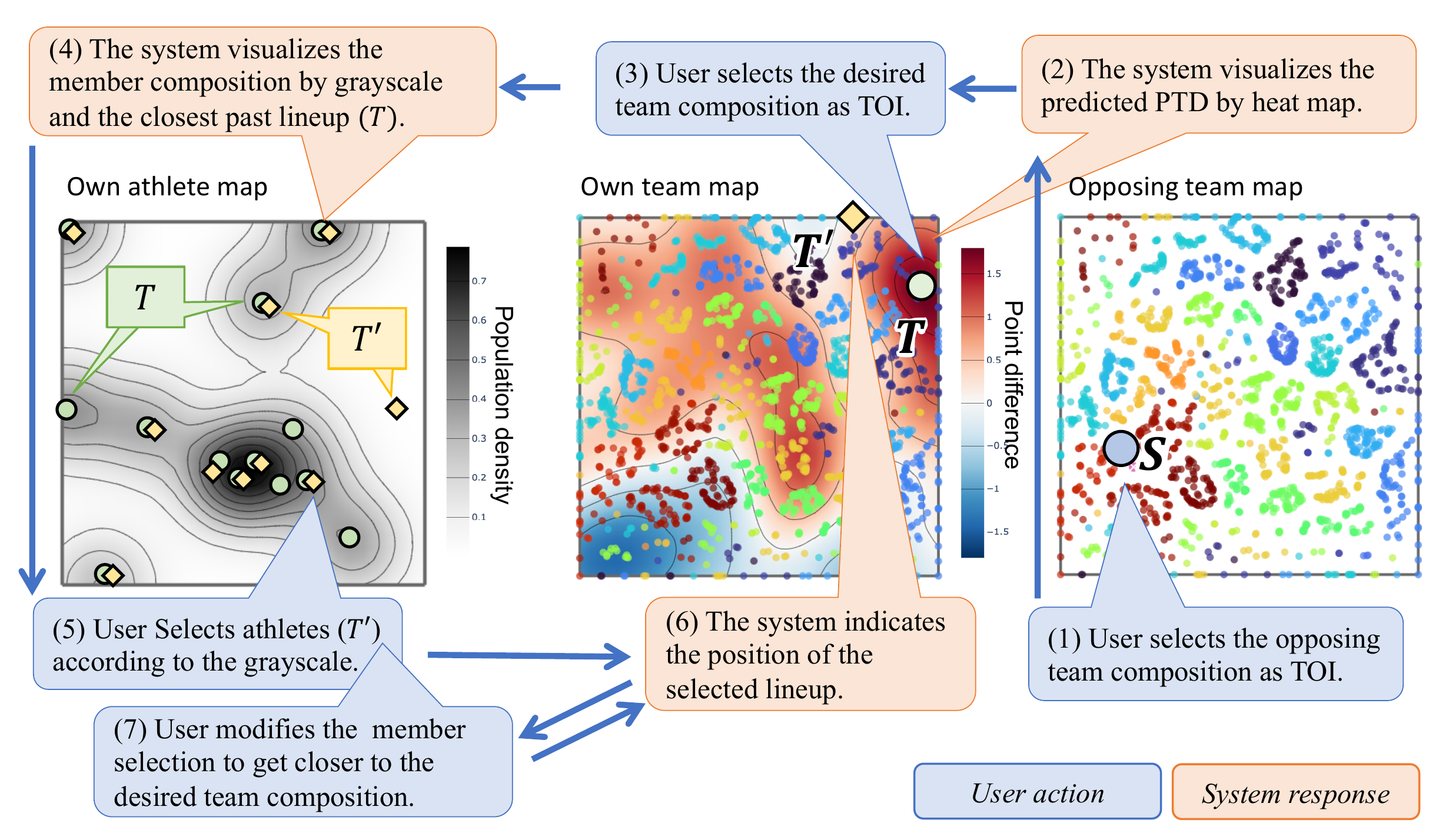}
    \caption{Example of the member selection process supported by the proposed VA system.}
    \figlabel{member_selection}
\end{figure*}    
\begin{figure*}
    \centering
    \includegraphics[width=0.85\textwidth]{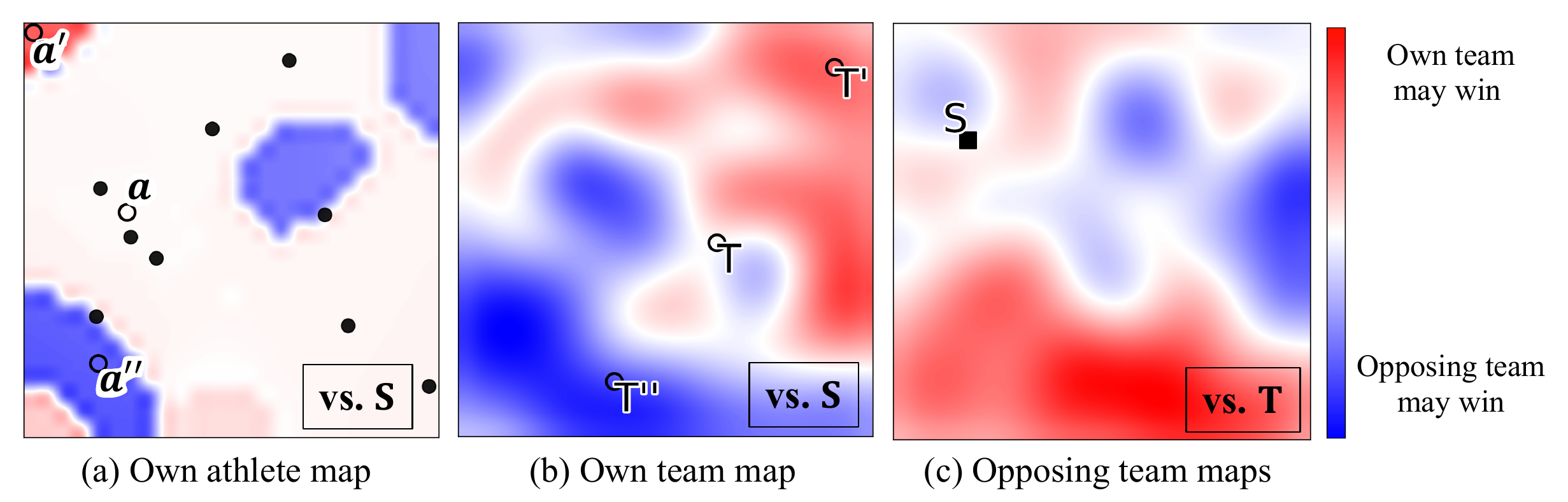}
    \caption{Table-top simulation of new lineups. Nine members were already selected (indicated by $\bullet$), and the 10th member was selected. (a) Predicted PTD with respect to the 10th member when the team plays a game with team $S$ in (c). $a$, $a'$, and $a''$ are the tentative candidates (indicated by $\circ$). (b) $T$, $T'$, and $T''$ indicate the position of the team when one of $a$, $a'$, and $a''$ is used, respectively. The color scale indicates the predicted PTD against team $S$. (c) The predicted PTD against team $T$.}
    \figlabel{result3}
\end{figure*}

\figref{member_selection} shows an example of the member selection process assisted by the VA system. In this scenario, suppose that the opposing team is located at $S$ on the opposing team map. If the user points to $S$, the system visualizes the predicted PTD on the own team map, thereby showing which own team may defeat the opposing team. Note that the user can view the predicted results of all teams on the team map simultaneously at a glance. By selecting a desired team (say, $T$ that shows the highest PTD) as the TOI, the system visualizes the member composition on the own athlete map as grayscale. Thus, by selecting athletes from the high-density region in the athlete map, the member composition becomes roughly equal to the desired composition. The system can also visualize the past lineup that is closest to $T$. After the user determines a lineup, the system indicates the location of the lineup on the own team map ($T'$). If necessary, the user can modify the lineup by trial and error while monitoring the lineup on the own team map. In this process, it is easy for the user to consider other factors, such as empirical knowledge, which are not built into the system. As shown in this example, the system allows a point on the team map to be transformed into the corresponding member composition in the athlete map, and vice versa. Such bidirectional mapping between teams and member compositions plays an essential role in member selection support. This is the remarkable achievement of the proposed method.

\figref{result3} shows another scenario in which the user already has a team consisting of nine members (indicated by black markers on the athlete map (a)), and tries to determine a new (i.e., 10th) member for the team. Additionally, suppose that the opposing team is located at $S$ on the opposing team map (c). In this scenario, the predicted PTD is visualized on the athlete map (a) as a function of the 10th member. Thus, the winning chance would be high/low if a member in the red/blue region is selected. Then, let $a$, $a'$, and $a''$ (white circles in the athlete map) be the candidates to be the 10th member. The corresponding teams $T$, $T'$, and $T''$, respectively, are indicated on the own team map (b). On the own team map, the predicted PTD is also visualized as a heat map. Thus, the winning chance is higher than 50\% for team $T'$, whereas it is less than 50\% for team $T''$. \figref{result3} (c) shows the predicted PTD when team $T$ plays a game against various opposing teams. It shows that team $T$ may defeat the opposing teams in the red region. In the above scenario, the task of the user is to select one of the candidates for the given team. The VA system can also be used in the opposite case, in which the manager needs to determine a suitable team for a newly added member. Thus, the VA system can support managers in assigning a new member to an existing team. 

Although only a few examples are provided above, they demonstrate the advantages of the proposed method. First, the system allows users to perform bidirectional exploration between a team property and member composition. For example, if a desired team is identified on the team map, the system visualizes the member composition on the athlete map. Conversely, if a lineup is provided, the system visualizes the position on the team map, as well as the predicted outcome. Second, the system allows users to indicate their TOI for each topographic map independently, which enables users to discover knowledge by unraveling the entangled factors. Third, the system allows users to perform analysis and prediction seamlessly. If a user selects an existing team or member as the TOI, the system shows the analysis result of past data, whereas the system predicts the result if the TOI did not exist in the past. Fourth, the system provides users with a bird's-eye view by providing topographic maps. Users can compare all team compositions in the team manifold simultaneously at a glance, and one-by-one comparison is not necessary. These four advantages correspond to the four requirements that we defined for the target system. In addition to these, the system has the advantages of the VA approach. Thus, it allows users to explore huge complex data interactively; it leaves the final decision to users; it allows users to make decisions together with considering their empirical knowledge or other factors that are not built in the system; and users can be accountable for their decisions. 

Note that the aim of this study is {\em not} to develop the demonstrated VA system; rather, our aim is to develop a general method for constructing such VA systems of set data. As already described, the proposed method allows developers to extend and adapt the system flexibly. Such plasticity is the advantage of the proposed approach as a system development method.

\section{Discussion} \label{sec:discussion}

\subsection{Evaluation and the five requirements}

Generally, the evaluation of a VA system is difficult, and there is no established protocol for evaluation \cite{Cui2019}. For an application-specific VA system, the system is usually evaluated using case studies, user studies, and expert reviews. If the study aims at developing a visual interface, then the operability is evaluated in subject experiments, whereas if the study aims at improving performance, then the system should be compared with other systems. By contrast, if the study aims at solving theoretical problems, the method should be evaluated logically to determine whether the problems are solved rather than using experiments. Because this study corresponds to the last case, we defined the five requirements as the criteria for evaluation. 

There may be a concern about the validity of these requirements. We can show that they are necessary conditions by considering what occurs if one or more requirements are not satisfied. In such a case, it is obvious that the system does not work as a VA system of set data anymore, or its ability is crucially limited. If either the first or fourth requirement is not achieved, the system cannot be used as a VA system of set data anymore. If the second requirement is not achieved, then the system can visualize only a limited aspect of the given data, and it cannot unravel the entangled factors anymore. If the third requirement is not achieved, the system cannot be used for decision-making for future cases. Clearly, each application requires additional requirements that are specific to the domain. To satisfy application-specific requirements, we highlight the importance of the fifth requirement. In this study, we focused on solving theoretical problems regarding handling set data. Evaluation using other criteria is our future task. Particularly, evaluation of operability and usefulness from the human-centered viewpoint is an important issue in the VA field.

\subsection{Extensions of the proposed method}

To apply the proposed system to practical cases, the VA system needs to be adapted according to cases; thus, plasticity is necessary. The key idea of this method is to model the generation process using the MNM, the structure of which is easy to adapt according to the data structure. In an implementation, the MNM can be estimated using combinations of GMM, where four connection styles are used consistently. Therefore, the proposed method can be adapted flexibly according to application cases. Because the MNM can be regarded as a Bayesian network model of manifolds, the method can be said to be universal rather than general. Although we have not programmed yet, it is possible to develop a software library so that the network structure can be extended incrementally on demand. Such on-demand plasticity would make the analytics process more dynamic.

This method also allows us to develop an integrated VA system by combining other analysis methods because topographic maps can be used as platforms on which various analysis results are visualized. For example, the results of PCA or factor analysis can be visualized on the maps. In this case, users only need to select those components as their TOI from the component list (i.e., the pull-down menu). Similarly, the outputs of other black-box systems can be visualized on the topographic maps.

For team formation support, it is easy to extend the method to cases in which the roles of members are given (e.g., positions in basketball, such as point guard and center). In this case, the team composition is represented by the joint probability of the member latent variable and role as $q(\mu,\rho\mid \tau)$, where $\rho$ represents the role. This extension would also be useful when the method is applied to a fashion outfit, in which the member and role are translated to a fashion item and category (e.g., shirts, pants, and coats) \cite{Li2017}. 

When new data are incrementally added over time, the proposed system can be updated by applying online learning. In this case, the members and teams are not fixed landmarks anymore; rather, they become `moving objects' in the topographic maps. For example, if the athlete stats are obtained for every month, then the stats vector and corresponding latent variable become time series $\vx_i(t)$ and $\mu_i(t)$, respectively. When we want to consider the latest performance of members, such an extension would be useful.

\section{Conclusion} \label{sec:conclusion}

In this paper, we proposed a method for VA systems of set data that satisfies five requirements. Because there are difficulties in handling set data, our proposed method is the first attempt to tackle these issues. The proposed method represents complex data using a network of manifold models, the structure of which can be flexibly extended according to the data structure. Therefore, we have proposed a general method for constructing VA systems of set data rather than a specific system for team management. 

In this work, we assumed that member selection support for team formation is a representative application, and demonstrated it by applying the proposed method to basketball team data. Although we still need to evaluate whether the method reaches a satisfactory level for team management, the concept of VA is more suited to such management issues than the conventional black-box approach. Therefore, we hope that this study will contribute not only to the VA field, but also the team management field as a methodology for a decision support system.

\section*{Acknowledgements}

This work was supported by JSPS KAKENHI [grant numbers 18K11472, 21K12061 and 20K19865] and ZOZO Research. We would like to thank Prof.\ D. Jahng from Kyushu Institute of Technology, Prof.\ H. Isogai from Kyushu Sangyo University, and Dr. T. Ohkubo from ZOZO Technologies for their valuable advice. We thank Mr.\ K. Senoura, who contributed to the preliminary stage of this study. 

\bibliographystyle{elsarticle-num}

\bibliography{refs}
\end{document}